\documentclass[twocolumn,trackchanges]{aastex701}

\usepackage{amsmath}
\usepackage{bm}

\usepackage{enumitem}

\begin{document}
\title{The 943~MHz radio–SFR relation of nearby galaxies observed with ASKAP}

\author[orcid=0009-0003-9052-1976]{Shengtao Wang}
\affiliation{School of Physics and Astronomy, Yunnan University, Kunming 650500, People's Republic of China}
\email[show]{wstfch@mail.ynu.edu.cn}  
\author[orcid=0000-0002-3464-5128]{Xiaohui Sun} 
\affiliation{School of Physics and Astronomy, Yunnan University, Kunming 650500, People's Republic of China}
\email[show]{xhsun@ynu.edu.cn}
\author[orcid=0000-0001-6239-3821]{Jiangtao Li}
\affiliation{Purple Mountain Observatory, Chinese Academy of Sciences, 10 Yuanhua Road, Nanjing 210023, People's Republic of China}
\email{pandataotao@gmail.com}
\author[orcid=0000-0002-9390-9672]{Chao-Wei Tsai}
\affiliation{National Astronomical Observatories, Chinese Academy of Sciences, 20A Datun Road, Beijing 100101, People's Republic of China}
\affiliation{Institute for Frontiers in Astronomy and Astrophysics, Beijing Normal University, Beijing 102206, People's Republic of China}
\affiliation{School of Astronomy and Space Science, University of Chinese Academy of Sciences, Beijing 100049, People's Republic of China}
\email{cwtsai@nao.cas.cn}
\author[0009-0006-3887-8988, gname=Jianghui, sname=Xu]{Jianghui Xu}
\affiliation{Department of Astronomy, University of Science and Technology of China, Hefei, Anhui 230026, People's Republic of China}
\affiliation{School of Astronomy and Space Science, University of Science and Technology of China, Hefei 230026, People's Republic of China}
\email{jianghuixu@mail.ustc.edu.cn}

\begin{abstract}

We investigate the integrated 943~MHz radio--infrared and radio--star-formation-rate (SFR) relations of 875 nearby galaxies within $D<50$~Mpc using public Australian SKA Pathfinder (ASKAP) observations. After excluding galaxies with identified or candidate AGN activity, 594 galaxies remain in the AGN-cleaned sample. We used Bayesian errors-in-variables models that account for measurement uncertainties, intrinsic scatter, and correlated distance errors. The AGN-cleaned radio--infrared relation is nearly linear, $L_{\nu,943}\propto L_{\nu,\mathrm{W4}}^{1.09\pm0.02}$, with an orthogonal intrinsic scatter of $0.21\pm0.01$~dex, and is tighter than the AGN-host relation. For 592 AGN-cleaned galaxies, the radio--SFR relation is approximately linear, $L_{\nu,943}\propto{\rm SFR}^{1.02\pm0.02}$, with an intrinsic scatter of $0.25\pm0.01$~dex. Semi-calorimetric models reveal positive secondary dependences on stellar mass, $\gamma=0.09\pm0.03$, and rotation velocity, $\delta=0.24\pm0.08$. The consistency between these two secondary trends suggests that the galaxy gravitational potential may weakly regulate CRE confinement and synchrotron-radiation efficiency. Our results provide an integrated 943~MHz radio--SFR calibration for nearby galaxies and show that, although radio luminosity scales nearly proportionally with SFR on average, weak secondary dependences on galaxy properties remain.

\end{abstract}

\keywords{\uat{Radio continuum emission}{1340} ---\uat{Cosmic rays}{329} ---\uat{Infrared astronomy}{786} ---\uat{Galaxy properties}{615} ---\uat{Star formation}{1569}}

\section{Introduction} 
\label{sec:intro}

Radio continuum emission provides a dust-unbiased tracer of star formation in galaxies. Near 1~GHz, it comprises thermal free--free emission from ionized gas and predominantly nonthermal synchrotron emission from cosmic-ray electrons (CREs) propagating through galactic magnetic fields. Because CREs are primarily accelerated in supernova remnants associated with short-lived massive stars, radio luminosity is closely linked to the recent star-formation rate (SFR)~\citep{Condon1992,Yun2001,Bell2003,Murphy2011}. Its insensitivity to dust extinction makes radio emission particularly valuable for studying dusty and highly inclined galaxies.

The connection between radio emission and star formation is further reflected in the tight radio--infrared correlation observed over several orders of magnitude in galaxy luminosity~\citep{Yun2001,Bell2003}. Infrared emission traces stellar radiation absorbed and re-emitted by dust, whereas low-frequency radio emission primarily traces synchrotron radiation from supernova-injected CREs. Although both are linked to recent star formation, neither is a strictly linear SFR tracer. Infrared luminosity depends on dust abundance, metallicity, temperature, geometry, and heating by older stellar populations~\citep{Kennicutt2009,Leroy2019}, while radio luminosity also depends on magnetic-field strength and the balance among CRE energy losses, transport, and escape~\citep{Condon1992,Lacki2010,Heesen2022}. These effects can produce nonlinear radio--infrared and radio--SFR relations. 

Under approximate energy equipartition between cosmic rays (CRs) and magnetic fields, enhanced gas turbulence can amplify magnetic fields through the small-scale dynamo, thereby increasing the synchrotron emissivity~\citep{Heesen2023}. In addition, CREs may escape more efficiently from small, low-mass galaxies, whereas larger and more massive systems can confine them for longer and allow a greater fraction of their energy to be radiated before escape. Galaxy mass, size, magnetic-field strength, and radio spectral index may therefore introduce secondary dependencies in the radio--SFR relation~\citep{Li2016,Heesen2022,Heesen2024,Heesen2026}.

Previous studies indicate that the radio--SFR relation depends on both observing frequency and spatial scale. Integrated measurements at gigahertz frequencies generally yield nearly linear or mildly superlinear relations, whereas steeper slopes are often found at $\rm \sim150~MHz$ frequencies~\citep{Heesen2014,Li2016,Wang2019,Heesen2022}. The stronger nonlinearity at low frequencies has been attributed in part to variations in CRE escape and calorimetric efficiency, including a dependence on galaxy mass~\citep{Smith2021,Heesen2022}. These effects can be described within a semi-calorimetric framework, in which CRE escape limits the fraction of injected energy emitted as synchrotron radiation. Variations in this efficiency are parameterized empirically through galaxy mass or other relevant properties~\citep{Heesen2022,Heesen2026}.

Spatially resolved studies provide a complementary view of the radio--SFR relation. Because CREs propagate away from their acceleration sites, radio emission is spatially smoothed relative to more localized SFR tracers~\citep[e.g.][]{Murphy2006,Berkhuijsen2013,Heesen2019,Heesen2024}. Resolved relations are therefore often sublinear even when the corresponding integrated relation is linear or superlinear. After accounting for CRE transport and calorimetric efficiency, however, the resolved relation can approach linearity~\citep{Heesen2024}. This contrast demonstrates that radio--SFR slopes cannot be interpreted solely in terms of calorimetry without considering CRE transport, spectral aging, and galaxy structure.

The radio--infrared and radio--SFR relations have been investigated over a broad frequency range, from the LOFAR bands near 144--150~MHz~\citep{Gurkan2018,Wang2019,Smith2021,Heesen2022,Heesen2024} to several gigahertz and above~\citep{Murphy2011,Li2016,Tabatabaei2017}. However, most empirical radio-based SFR calibrations were established near 1.4~GHz~\citep{Condon1992,Yun2001,Bell2003,Hopkins2003,Davies2017}, and comparatively few large, homogeneous studies have characterized the integrated relations of nearby galaxies specifically near 1~GHz. At this frequency, synchrotron emission remains dominant, while free--free absorption and low-energy ionization losses are less important than at meter wavelengths.

The Evolutionary Map of the Universe (EMU) survey conducted with the Australian SKA Pathfinder (ASKAP) provides sensitive, wide-area radio continuum imaging of the southern sky near 943~MHz~\citep{Norris2011,Norris2021}. These observations therefore offer a valuable intermediate-frequency view between the extensively studied LOFAR and conventional 1.4~GHz regimes. 

In this work, we use publicly released ASKAP/EMU data to investigate the integrated 943~MHz radio properties of galaxies within $D<50$~Mpc. We characterize the relation between 943~MHz radio luminosity and Wide-field Infrared Survey Explorer (WISE) W4 ($\rm 22\,\mu m$) luminosity, derive the integrated radio--SFR relation for the AGN-cleaned sample, and use
semi-calorimetric models to test for secondary dependences on stellar
mass and rotation velocity. Our goal is to determine whether radio luminosity near 1~GHz provides a universal SFR calibration or is modulated by the galaxy gravitational potential through variations in CRE confinement and synchrotron-radiation efficiency.

This paper is organized as follows. Section~\ref{sec:data} describes the sample selection, integrated flux measurements at 943~MHz, AGN classification, and ancillary WISE photometry, stellar-mass, SFRs, and galaxy rotation velocities data. Section~\ref{sec:method} presents the luminosity calculations and Bayesian regression methodology. The radio--infrared, radio--SFR, and semi-calorimetric relations involving stellar mass and rotation velocity are presented in Section~\ref{sec:results} and discussed in Section~\ref{sec:discussion}. The main conclusions are summarized in Section~\ref{sec:conclusions}.

\section{Data}
\label{sec:data}

\subsection{Sample selection}
\label{sec:sample_selection}

We constructed the parent sample by combining the 50~Mpc Galaxy Catalog (50MGC; \citealt{Ohlson2024}), containing 15,424 galaxies, with the $z=0$ Multiwavelength Galaxy Synthesis catalog (z0MGS; \citealt{Leroy2019}), containing 15,748 galaxies. After cross-matching and removing duplicate entries, the combined catalog contained 23,787 distinct galaxies within $D<50$~Mpc.

\begin{figure} \includegraphics[width=\columnwidth]{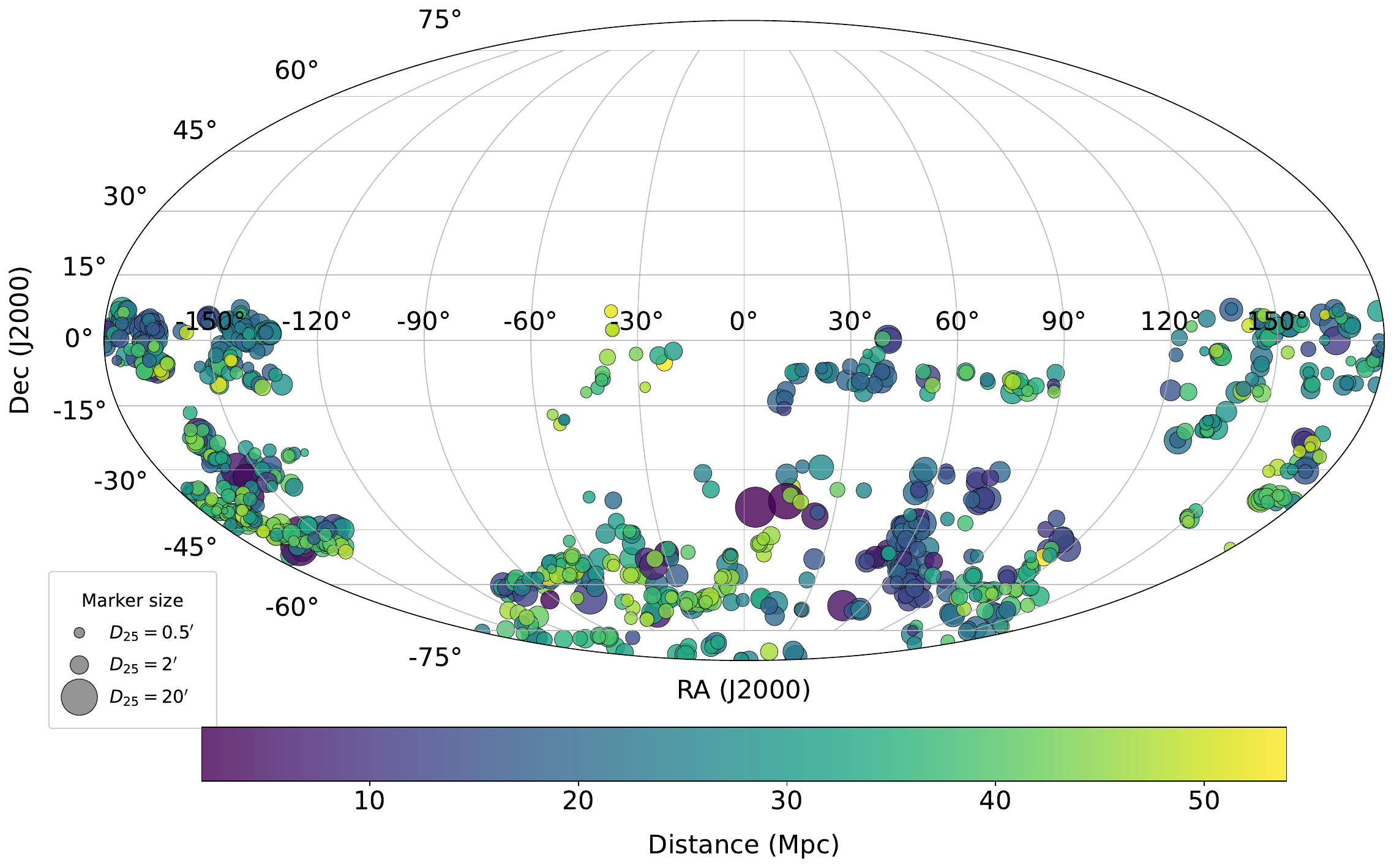}
    \caption{Sky distribution of the 875 galaxies in the sample. The marker color indicates the adopted distance, while the marker area scales with the optical diameter $D_{25}$.} 
\label{fig:gal_sample}
\end{figure}

The combined catalog was then restricted to the ASKAP/EMU-accessible sky by requiring $\delta<+30\degr$~\citep{Hopkins2025}, while galaxies close to the Galactic plane were excluded using $|b|>10\degr$ to reduce contamination from Galactic foreground emission and dust extinction. These cuts yielded a parent sample of 17,188 galaxies. For each remaining galaxy, we searched the CSIRO ASKAP Science Data Archive \textit \rm {(CASDA)}\footnote{\url{https://data.csiro.au/domain/casdaObservation}} for publicly available ASKAP/EMU 943~MHz Stokes-$I$ images and downloaded a cutout centered on the catalog position when a suitable image was available. We retained only galaxies with a peak surface brightness of $I_{\nu,\mathrm{peak}}>500~\mu{\rm Jy~beam^{-1}}$, corresponding to at least a $10\sigma$ detection for the typical rms noise level of
$30$--$50~\mu{\rm Jy~beam^{-1}}$ in the 10-h EMU observations, which have a typical synthesized beam size of $15\arcsec$. This selection yielded 875 radio-selected galaxies, representing approximately $5.1\%$ of the EMU-accessible parent sample. Their sky distribution is show in Figure~\ref{fig:gal_sample}.

\subsection{Integrated 943~MHz flux-density measurements}
\label{sec:radio_flux_measurement}

Integrated 943~MHz flux densities were measured from the ASKAP/EMU cutouts using an automated Python procedure. The background, $I_{\rm bg}$, and rms noise, $\sigma_{\rm rms}$, were estimated from the finite image pixels using iterative $3\sigma$-clipped statistics, adopting the clipped median and standard deviation, respectively. Candidate source pixels were identified by applying a surface-brightness threshold to the total-intensity image,
\begin{equation}
    I_{\nu}(x,y) > I_{\rm bg} + 3\sigma_{\rm rms},
    \label{eq:radio_detection_threshold}
\end{equation}
Contiguous pixels satisfying Eq.~(\ref{eq:radio_detection_threshold}) were grouped into connected emission islands. 

Before calculating the integrated flux density, the estimated background level was subtracted from every pixel in the selected source region. The integrated flux density was therefore calculated as
\begin{equation}
    S_{\nu}
    =
    \frac{1}{N_{\rm pix,beam}}
    \sum_{i\in\mathcal{A}}
    \left(I_{\nu,i}-I_{\rm bg}\right),
    \label{eq:integrated_radio_flux}
\end{equation}
where $\mathcal{A}$ denotes the selected source mask, and $N_{\rm pix,beam}=\Omega_{\rm beam}/\Omega_{\rm pix}$ is the number of image pixels per synthesized beam. Here, $\Omega_{\rm beam}$ and $\Omega_{\rm pix}$ are the areas of the synthesized beam and an individual image pixel, respectively.

The uncertainty on the integrated flux density includes both the thermal noise term and a fractional flux-calibration term. The number of independent beams in the aperture is $N_{\rm beam}=N_{\rm pix}/N_{\rm pix,beam}$, where $N_{\rm pix}$ is the number of pixels in the target mask. The thermal uncertainty is therefore $\sigma_{\rm th}=\sigma_{\rm rms}\sqrt{N_{\rm beam}}$. We adopted a fractional ASKAP flux-calibration uncertainty of 5\%, giving $\sigma_{\rm cal}=0.05\,S_{943}$. The final uncertainty on the integrated 943 MHz flux density was obtained by adding these terms in quadrature:
\begin{equation}
    \sigma_{S_{943}}
    =
    \sqrt{
    \sigma_{\rm th}^{2}
    +
    \sigma_{\rm cal}^{2}
    }.
\end{equation}

\subsection{AGN host galaxy catalog}
\label{subsec:agn_hosts}

AGN hosts were identified by combining catalog cross-matching, literature-based object classifications, and mid-infrared color diagnostics. First, every galaxy in the parent sample was independently cross-matched, using a radius of $5\arcsec$, against the second data release of the BAT AGN Spectroscopic Survey (BASS DR2; \citealt{Koss2022}), the Swift/BAT 105-month catalog~\citep{Oh2018}, and version 8 of the Million Quasars catalog (Milliquas v8; \citealt{Flesch2023}). A galaxy was assigned an AGN flag when it was associated with an entry in any one of these catalogs.

As a complementary literature-based search, SIMBAD was queried by object name~\citep{Wenger2000}. Galaxies classified as an active galactic nucleus (\texttt{AGN}), a Seyfert galaxy (\texttt{SyG}), a Type~1 or Type~2 Seyfert galaxy (\texttt{Sy1} or \texttt{Sy2}), a BL~Lacertae object (\texttt{BLL}), a blazar (\texttt{Bla}), a quasi-stellar object (\texttt{QSO}), or a LINER (\texttt{LIN}) were assigned an AGN-related flag.

WISE mid-infrared colors were then used as an additional diagnostic for galaxies without an AGN identification from the external catalogs or SIMBAD. Using Vega-system magnitudes, we adopted the Stern criteria, $W1-W2\geq0.8$~\citep{Stern2012}, and the Jarrett AGN wedge, defined by $2.2<W2-W3<4.2$ and $0.1(W2-W3)+0.38<W1-W2<1.7$~\citep{Jarrett2011}. A source identified solely by the strong Stern color criterion was included in the
\texttt{Other AGN} category. This category also contains sources with an external or SIMBAD AGN classification that could not be assigned to the \texttt{Type 1 AGN}, \texttt{Type 2 AGN}, or \texttt{LINER} categories. Sources with valid WISE colors that satisfied neither the Stern nor the
Jarrett AGN criterion, but also fell outside our adopted star-forming-like region, $W1-W2<0.5$ and $W2-W3>1.5$, were classified as \texttt{WISE\_candidate}. This designation indicates an ambiguous
mid-infrared classification rather than a secure AGN identification. Nevertheless, these sources were conservatively excluded from the AGN-cleaned sample.

The final catalog contains 875 galaxies, of which 281 carried an AGN-related flag. Among these, 235 had AGN-related classifications from the external catalogs or SIMBAD, one additional galaxy was identified solely through the strong WISE AGN color criterion, and 45 were classified as \texttt{WISE\_candidate}. Based on the representative AGN classification, the AGN-related sample comprises 59 \texttt{Type 1 AGN}, 147 \texttt{Type 2 AGN}, 10 \texttt{LINER}, 45 \texttt{WISE\_candidate}, and 20 \texttt{Other AGN} sources. The remaining 594 galaxies had no AGN identification from the adopted diagnostics and constitute our AGN-cleaned sample.

\subsection{WISE photometry, stellar masses, and star-formation rates}
\label{sec:wise_sfr_mass}

For 849 of the 875 galaxies, the WISE photometry, stellar masses, and star-formation rates (SFRs) were obtained from the z0MGS catalog~\citep{Leroy2019}. The z0MGS atlas combines Galaxy Evolution Explorer (GALEX) far-ultraviolet (FUV) and near-ultraviolet (NUV) observations with Wide-field Infrared Survey Explorer (WISE) imaging at 3.4, 4.6, 12, and 22~$\mu{\rm m}$ (W1--W4). We adopted the catalog-integrated WISE flux densities in Jy and their corresponding uncertainties.

Stellar masses in z0MGS were derived from the WISE W1 luminosity using a mass-to-light ratio that varies with galaxy properties. We adopted the resulting catalog values of $\log_{10}(M_\star/M_\odot)$, where $M_\star$ denotes the stellar mass, together with their uncertainties.

The z0MGS SFR estimates combine the unobscured star formation traced by GALEX UV emission with the dust-obscured component traced by WISE W4:
\begin{equation}
    {\rm SFR}
    =
    C_{\rm UV}\,\nu L_{\nu,{\rm UV}}
    +
    C_{\rm W4}\,\nu L_{\nu,{\rm W4}},
    \label{eq:z0mgs_sfr}
\end{equation}
where the calibration coefficients were tied to the population-synthesis
SFRs of \citet{Salim2016,Salim2018}. The FUV+W4 estimator was used when
FUV measurements were available, NUV+W4 was used otherwise, and W4 alone
was adopted when no reliable GALEX measurement was available. We used
the catalog values of
\(\log_{10}[{\rm SFR}/(M_\odot\,{\rm yr}^{-1})]\) and their associated
uncertainties.

The remaining 26 galaxies are included in the 50MGC but have no counterparts in z0MGS. For these galaxies, WISE image cutouts were retrieved from the unWISE AllWISE coadds~\citep{Lang2014}, and integrated flux densities were measured directly from the images. The local background and image noise were estimated following the same procedure used for the 943~MHz measurements in Section~\ref{sec:radio_flux_measurement}. The flux-density uncertainties include contributions from both image noise and an adopted 5\% absolute flux-calibration uncertainty.

We adopted the W4-only SFR estimator for all 26 galaxies:
\begin{equation}
    {\rm SFR}
    =
    10^{-42.63}
    \left(
        \frac{\nu L_{\nu,{\rm W4}}}
             {{\rm erg\,s^{-1}}}
    \right)
    M_\odot\,{\rm yr}^{-1},
    \label{eq:w4_only_sfr}
\end{equation}
following the z0MGS calibration of~\citet{Leroy2019}. The SFR uncertainty was calculated by propagating the W4 flux-density uncertainty and adding a 0.2-dex calibration uncertainty in quadrature.

For 16 of these 26 galaxies, the stellar masses and their uncertainties were available directly from 50MGC and were adopted without recalculation. For the remaining 10 galaxies, the stellar mass was estimated from the W1 luminosity using the z0MGS W4-to-W1 mass-to-light-ratio prescription. Defining $Q=\mathrm{log}_{10}(L_{\rm W4}/L_{\rm W1})$, the adopted W1 stellar mass-to-light ratio was~\citep{Leroy2019}
\begin{equation}
    \Upsilon_\star^{3.4}
    =
    \begin{cases}
        0.5, & Q \leq 0,\\
        0.5-0.4Q, & 0<Q<0.75,\\
        0.2, & Q \geq 0.75.
    \end{cases}
    \label{eq:w4w1_mtol}
\end{equation}
The stellar mass was then calculated as $M_\star=\Upsilon_\star^{3.4}L_{\rm W1}$. Its uncertainty was determined by propagating the W1 and W4 flux-density uncertainties using Monte Carlo sampling and adding a 0.1-dex systematic uncertainty in the mass-to-light ratio.

In the final catalog, the adopted SFR estimator was FUV+W4 for 564 galaxies, NUV+W4 for 50 galaxies, and W4-only for 261 galaxies. Eight galaxies have negative background-subtracted W4 flux densities and were excluded from the radio--W4 analysis. Four of these belong to the AGN-cleaned sample. Two of these four retain finite FUV+W4 SFR estimates because their FUV components remain positive and dominant, and were therefore included in the radio--SFR analysis. The other two have non-finite SFR estimates and were excluded, leaving 592 galaxies for the radio--SFR fit.

\subsection{Galaxy rotation velocities}
\label{sec:rotation_velocity}

Galaxy rotation velocities were obtained from the \textit \rm {(HyperLEDA)}\footnote{\url{http://atlas.obs-hp.fr/hyperleda/}}  database~\citep{Makarov2014}. We adopted the HyperLEDA parameter $v_{\rm rot}$, which represents the homogenized maximum gas rotation velocity corrected for galaxy inclination. These velocities are mainly derived from H\,{\sc i} line-width and resolved rotation-curve measurements.

Rotation velocities were available for 648 of the 875 galaxies. The cataloged HyperLEDA uncertainties were adopted for 514 galaxies. For the remaining 134 galaxies with a valid rotation velocity but no reported uncertainty, we conservatively assigned a fractional uncertainty of 10\%. After removing galaxies with identified or candidate AGN activity, 454 galaxies with valid rotation-velocity measurements remained in the AGN-cleaned sample.

\section{Method}
\label{sec:method}
All scaling relations were analyzed within a Bayesian errors-in-variables framework, accounting for measurement uncertainties and intrinsic astrophysical scatter. A symmetric orthogonal likelihood was adopted for the radio--infrared luminosity relation because neither luminosity has a preferred role as the independent variable. The radio--SFR and semi-calorimetric relations were instead fitted conditionally, with radio luminosity treated as the dependent quantity and the uncertainties in all predictors propagated into the residual variance.

The monochromatic spectral luminosity was calculated as
\begin{equation}
    L_{\nu}
    =
    4\pi D^{2}S_{\nu}\times10^{-26},
    \label{eq:spectral_luminosity}
\end{equation}
where $D$ is the galaxy distance in meters and $S_{\nu}$ is the flux density in Jy, giving $L_{\nu}$ in ${\rm W\,Hz^{-1}}$. The $K$-correction was neglected because the galaxies in our sample have very low redshifts, $z\lesssim0.01$. Its logarithmic uncertainty is
\begin{equation}
    \sigma_{\log L_{\nu}}^{2}
    =
    \left(
        \frac{\sigma_{S_{\nu}}}
             {S_{\nu}\ln 10}
    \right)^{2}
    +
    \left(2\sigma_{\log D}\right)^{2},
    \label{eq:luminosity_uncertainty}
\end{equation}
where $\sigma_{S_{\nu}}$ is the flux-density uncertainty and $\sigma_{\log D}$ is the distance uncertainty in dex. In the likelihood, the flux-density and distance terms were retained separately. When two luminosities were calculated using the same galaxy distance, the shared distance uncertainty was treated as covariance rather than as two independent error contributions.

\subsection{Radio--infrared luminosity regression}
The 943~MHz and W4 luminosities were fitted using the Bayesian orthogonal errors-in-variables model introduced above. We defined
\begin{equation}
    w_{i}
    =    \log_{10}L_{\nu,W4,i}
    \qquad
    r_{i}
    =    \log_{10}L_{\nu,943,i},
\end{equation}
and fitted the linear relation $r_i=aw_i+b$.    

The likelihood was constructed from the perpendicular distance of each observed point from the fitted line. The measurement uncertainties in both luminosities were projected onto the perpendicular direction, together with their covariance arising from the shared $D^{2}$ dependence. An additional orthogonal intrinsic-scatter term, $\sigma_{\perp,\mathrm{int}}$, was included to describe astrophysical dispersion beyond the measurement uncertainties. 

\subsection{Radio--SFR regression}

The radio--SFR relation was fitted in logarithmic space as
\begin{equation}
    r_i
    =
    a s_i+b+\epsilon_i,
    \qquad
    \epsilon_i
    \sim
    \mathcal{N}(0,\sigma_{\rm int}^{2}),
    \label{eq:radio_sfr_model}
\end{equation}
where $s_i=\log_{10}{\rm SFR}_i$, $a$ is the radio--SFR slope, and $b$ is the normalization. Here, $\mathcal{N}$ denotes a normal distribution with zero mean and standard deviation $\sigma_{\rm int}$, which describes the intrinsic scatter in radio luminosity at fixed SFR.

Measurement uncertainties in both variables, together with the shared distance uncertainty and intrinsic scatter, were propagated into the effective residual variance,
\begin{equation}
    \sigma_{{\rm eff},i}^{2}
    =
    \sigma_{\log S_{943},i}^{2}
    +
    a^{2}\sigma_{s,i}^{2}
    +
    (1-a)^{2}\sigma_{D,i}^{2}
    +
    \sigma_{\rm int}^{2},
    \label{eq:radio_sfr_effective_variance}
\end{equation}
where $\sigma_{\log S_{943},i}$ is the logarithmic uncertainty associated with the 943\,MHz flux density, $\sigma_{s,i}$ is the cataloged uncertainty in $\log_{10}{\rm SFR}$ excluding the separately propagated distance term, and $\sigma_{D,i}=2\sigma_{\log D,i}$.

Because both SFR and radio luminosity scale as $D^{2}$, their distance errors are correlated. The coefficient $(1-a)$ accounts for this shared distance contribution in the residual $r_i-(as_i+b)$. The Gaussian log-likelihood was
\begin{equation}
    \ln\mathcal{L}
    =
    -\frac{1}{2}
    \sum_i
    \left[
        \frac{
            \left[r_i-(as_i+b)\right]^{2}
        }{
            \sigma_{{\rm eff},i}^{2}
        }
        +
        \ln\left(
            2\pi\sigma_{{\rm eff},i}^{2}
        \right)
    \right].
    \label{eq:conditional_log_likelihood}
\end{equation}
The slope, normalization, and intrinsic scatter were inferred
simultaneously.

\subsection{Semi-calorimetric models}

To test whether radio luminosity has a secondary dependence on stellar mass, we fitted
\begin{equation}
\begin{split}
    r_i
    ={}&
    \log_{10}L_{\rm C}
    +
    a s_i
    +
    \gamma m_i
    +
    \epsilon_i,
\end{split}
\label{eq:mass_semical_model}
\end{equation}
where
\begin{equation}
    m_i
    =
    \log_{10}
    \left(
        \frac{M_{\star,i}}{10^{10}M_\odot}
    \right).
\end{equation}
The coefficient $a$ describes the dependence of radio luminosity on SFR, while $\gamma$ quantifies the additional stellar-mass dependence. The effective variance was
\begin{equation}
\begin{split}
    \sigma_{{\rm eff},i}^{2}
    ={}&
    \sigma_{\log S_{943},i}^{2}
    +
    a^{2}\sigma_{s,i}^{2}
    +
    \gamma^{2}\sigma_{m,i}^{2}
    \\
    &+
    (1-a-\gamma)^{2}\sigma_{D,i}^{2}
    +
    \sigma_{\rm int}^{2}.
\end{split}
\label{eq:mass_semical_effective_variance}
\end{equation}
Here, $\sigma_{m,i}$ is the cataloged uncertainty in $m_i=\log_{10}(M_{\star,i}/10^{10}M_\odot)$, excluding the separately propagated distance term. The factor $(1-a-\gamma)$ accounts for the shared $D^2$ dependence of the radio luminosity, SFR, and stellar mass.

An analogous model was used to test the dependence on rotation velocity,
\begin{equation}
    r_i
    =
    \log_{10}L_{\rm C}
    +
    a s_i
    +
    \delta v_i
    +
    \epsilon_i,
    \label{eq:vrot_semical_model}
\end{equation}
where
\begin{equation}
\begin{aligned}
    v_i
    &=
    \log_{10}
    \left(
        \frac{V_{{\rm rot},i}}
             {200\,{\rm km\,s^{-1}}}
    \right), \\
    \sigma_{v,i}
    &=
    \frac{1}{\ln 10}
    \frac{\sigma_{V_{{\rm rot},i}}}
         {V_{{\rm rot},i}}.
\end{aligned}
\end{equation}
The corresponding effective variance was
\begin{equation}
\begin{split}
    \sigma_{{\rm eff},i}^{2}
    ={}&
    \sigma_{\log S_{943},i}^{2}
    +
    a^{2}\sigma_{s,i}^{2}
    +
    \delta^{2}\sigma_{v,i}^{2}
    \\
    &+
    (1-a)^{2}\sigma_{D,i}^{2}
    +
    \sigma_{\rm int}^{2}.
\end{split}
\label{eq:vrot_semical_effective_variance}
\end{equation}
Rotation velocity does not depend on the adopted galaxy distance; therefore, it does not contribute to the shared-distance covariance. In both semi-calorimetric models, $a$, the secondary dependence ($\gamma$ or $\delta$), the normalization, and the intrinsic scatter were fitted simultaneously using the likelihood in Eq.~(\ref{eq:conditional_log_likelihood}).

Posterior distributions were sampled using the affine-invariant MCMC sampler \texttt{emcee}~\citep{ForemanMackey2013}, with broad uniform priors on the sampled parameters. The radio--infrared regressions used 64 walkers. The radio--SFR regression and the semi-calorimetric models each used 80 walkers. All walkers were evolved for 5000 steps. The first 1500 steps were discarded as burn-in, and the remaining chains were thinned by a factor of 10. Reported parameters are posterior medians, with uncertainties defined by the 16th and 84th percentiles. The plotted $3\sigma$ intervals correspond to the pointwise 0.135th and 99.865th percentiles. Fit-uncertainty regions represent uncertainty in the mean relation, whereas posterior-predictive regions additionally include the inferred intrinsic scatter.

\begin{figure*}[!htbp]
    \centering
    \includegraphics[width=0.47\textwidth]{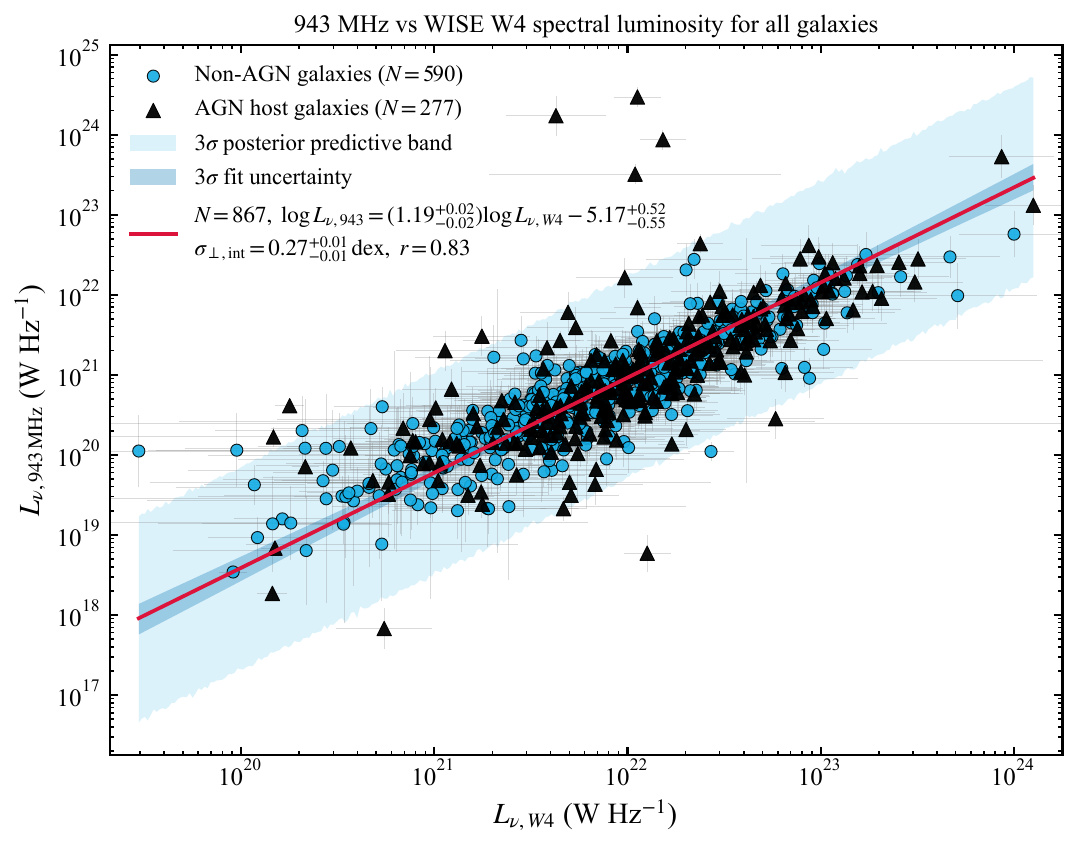}
    \includegraphics[width=0.47\textwidth]{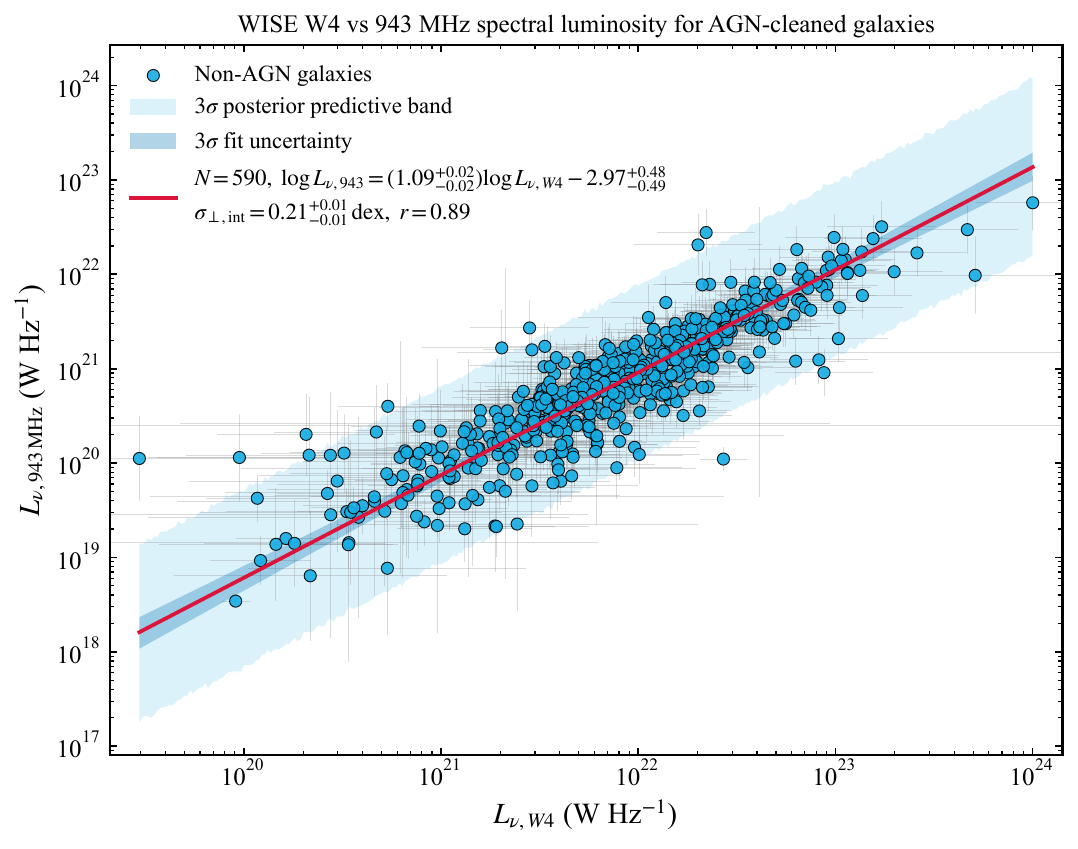}
    \caption{Relation between the WISE W4 spectral luminosity and the 943~MHz radio luminosity. The left panel shows the full sample of galaxies with valid measurements ($N=867$), while the right panel shows the AGN-cleaned sample ($N=590$). Blue circles denote galaxies with no identified AGN, whereas black triangles indicate AGN-host galaxies. The red lines represent the median \texttt{emcee} regression results. The dark-blue regions show the $3\sigma$ uncertainty in the fitted mean relation, while the light-blue regions indicate the corresponding $3\sigma$ posterior-predictive intervals, including the inferred intrinsic scatter.}
    \label{fig:radio_ir_relation}
\end{figure*}

\section{Results}
\label{sec:results}

\subsection{Radio--infrared luminosity relation}
\label{sec:results_radio_ir}

Figure~\ref{fig:radio_ir_relation} shows the relation between the 943~MHz radio luminosity and the WISE W4 spectral luminosity. The left panel presents the full sample, including both AGN-cleaned galaxies and AGN hosts, while the right panel shows the AGN-cleaned sample. We fitted both samples in logarithmic space using
\begin{equation}
    \log_{10}
    \left(
        \frac{L_{\nu,943}}{{\rm W\,Hz^{-1}}}
    \right)
    =
    a
    \log_{10}
    \left(
        \frac{L_{\nu,W4}}{{\rm W\,Hz^{-1}}}
    \right)
    + b.
    \label{eq:result_radio_ir}
\end{equation}

The full sample contains 867 galaxies with valid W4 and 943~MHz luminosities and associated uncertainties. Using the Bayesian orthogonal errors-in-variables regression described in Section~\ref{sec:method}, we obtain
\begin{equation}
\begin{aligned}
    \log_{10}
    \left(
        \frac{L_{\nu,943}}{{\rm W\,Hz^{-1}}}
    \right)
    ={}&
    \left(
        1.19_{-0.02}^{+0.02}
    \right)
    \log_{10}
    \left(
        \frac{L_{\nu,W4}}{{\rm W\,Hz^{-1}}}
    \right)
    \\
    &-
    \left(
        5.17_{-0.55}^{+0.52}
    \right).
\end{aligned}
\label{eq:result_radio_ir_all}
\end{equation}
The relation has an intrinsic scatter of $\sigma_{\perp,\rm int}=0.27_{-0.01}^{+0.01}$ dex, and a Pearson correlation coefficient of $r=0.83$. The two luminosities are therefore strongly correlated.

For the AGN-cleaned sample, 590 galaxies have valid measurements and uncertainties. The fitted relation is
\begin{equation}
\begin{aligned}
    \log_{10}
    \left(
        \frac{L_{\nu,943}}{{\rm W\,Hz^{-1}}}
    \right)
    ={}&
    \left(
        1.09_{-0.02}^{+0.02}
    \right)
    \log_{10}
    \left(
        \frac{L_{\nu,W4}}{{\rm W\,Hz^{-1}}}
    \right)
    \\
    &-
    \left(
        2.97_{-0.49}^{+0.48}
    \right).
\end{aligned}
\label{eq:result_radio_ir_sf}
\end{equation}
The intrinsic scatter decreases to $\sigma_{\perp,\rm int}=0.21_{-0.01}^{+0.01}$~dex, while the Pearson correlation coefficient increases to $r=0.89$.

\begin{figure} \includegraphics[width=\columnwidth]{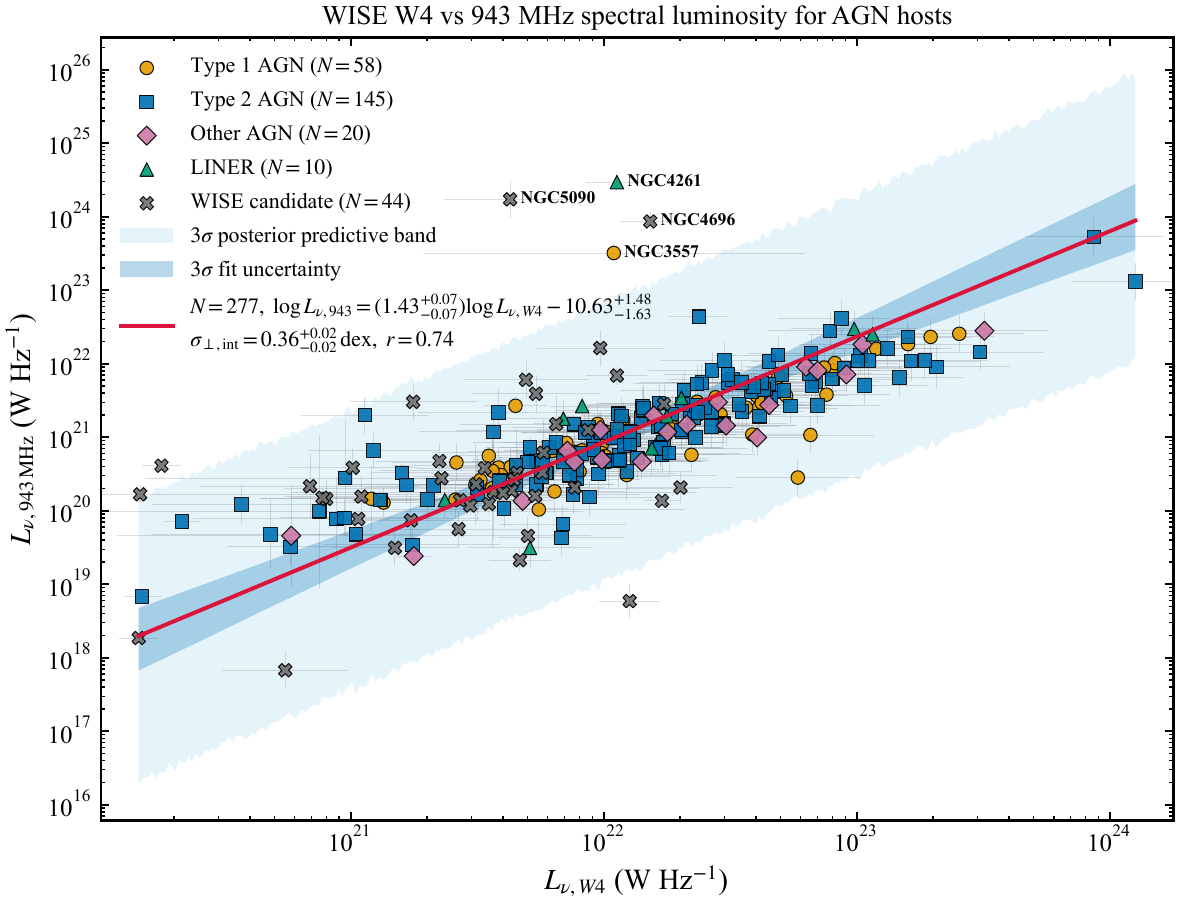}
    \caption{Relation between the WISE W4 spectral luminosity and the 943~MHz radio spectral luminosity for AGN-host galaxies. The 277 galaxies with valid luminosity and uncertainty measurements are distinguished according to their representative AGN classifications using different colors and symbols. The four galaxies with the largest positive offsets from the fitted relation, NGC~5090, NGC~4261, NGC~4696, and NGC~3557, are labelled. The red line, dark- and light-blue regions have the same meanings as in Figure~\ref{fig:radio_ir_relation}.} 
\label{fig:agn_w4_943}
\end{figure}

For comparison, we also fitted the radio--infrared relation for the AGN-host sample, as shown in Figure~\ref{fig:agn_w4_943}. The different AGN categories are distinguished by different colors and symbols. The resulting relation is
\begin{equation}
\begin{aligned}
    \log_{10}
    \left(
        \frac{L_{\nu,943}}{{\rm W\,Hz^{-1}}}
    \right)
    ={}&
    \left(
        1.43_{-0.07}^{+0.07}
    \right)
    \log_{10}
    \left(
        \frac{L_{\nu,W4}}{{\rm W\,Hz^{-1}}}
    \right)
    \\
    &-
    \left(
        10.63_{-1.63}^{+1.48}
    \right).
\end{aligned}
\label{eq:result_radio_ir_agn}
\end{equation}
The AGN-host relation has an intrinsic scatter of $\sigma_{\perp,\rm int}=0.36_{-0.02}^{+0.02}$~dex, and a Pearson correlation coefficient of $r=0.74$. 

As a sensitivity test, we repeated the fit after excluding the four galaxies with the largest positive offsets from the relation: NGC~5090, NGC~4261, NGC~4696, and NGC~3557. For the remaining 273 AGN hosts, the slope decreases to $1.26_{-0.05}^{+0.05}$ and the orthogonal intrinsic scatter decreases to $\sigma_{\perp,\rm int}=0.29_{-0.01}^{+0.01}$~dex, while the Pearson correlation coefficient increases to $r=0.83$. Thus, these radio-excess sources contribute substantially to the steep slope and large dispersion of the AGN-host relation. Nevertheless, even after their exclusion, the AGN-host relation remains significantly steeper and more dispersed than the AGN-cleaned relation. The slope and intrinsic scatter of the full-sample relation lie between those obtained for the AGN-cleaned and AGN-host samples. The physical origin of these differences is discussed in Section~\ref{sec:discussion_radio_ir}.
   
\subsection{Radio--SFR relation}
\label{sec:results_radio_sfr}

\begin{figure} \includegraphics[width=\columnwidth]{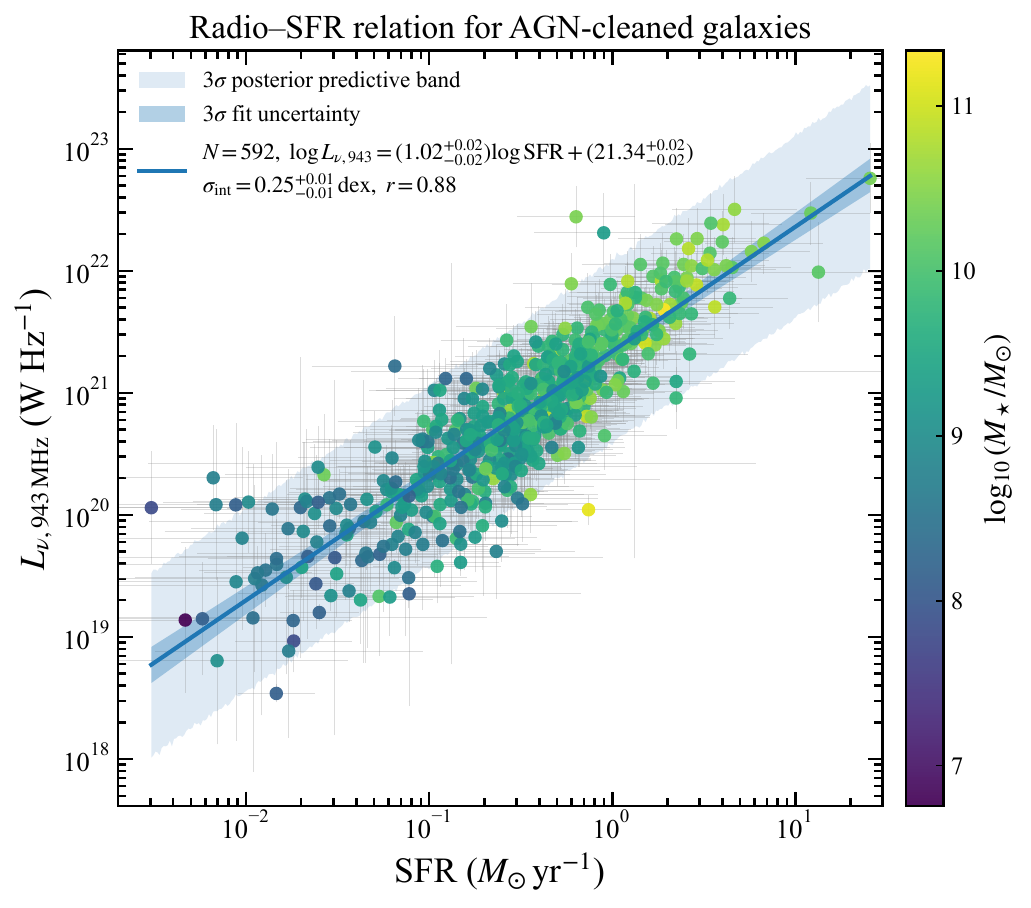}
    \caption{The 943~MHz radio luminosity as a function of star-formation rate for the AGN-cleaned galaxy sample ($N=592$). The data points are color-coded by total stellar mass. The red line and the dark- and light-blue regions have the same meanings as in Figure~\ref{fig:radio_ir_relation}.} 
\label{fig:radio_sfr}
\end{figure}
We fitted a power-law relation to the AGN-cleaned sample using the Bayesian conditional errors-in-variables regression described in Section~\ref{sec:method}:
\begin{equation}
    \log_{10}
    \left(
        \frac{L_{\nu,943}}{{\rm W\,Hz^{-1}}}
    \right)
    =
    a
    \log_{10}
    \left(
        \frac{\mathrm{SFR}}
             {M_{\odot}\,{\rm yr}^{-1}}
    \right)
    + b.
    \label{eq:result_radio_sfr_powerlaw}
\end{equation}
The resulting relation is
\begin{equation}
\begin{aligned}
    \log_{10}
    \left(
        \frac{L_{\nu,943}}{{\rm W\,Hz^{-1}}}
    \right)
    ={}&
    \left(
        1.02_{-0.02}^{+0.02}
    \right)
    \log_{10}
    \left(
        \frac{\mathrm{SFR}}
             {M_{\odot}\,{\rm yr}^{-1}}
    \right)
    \\
    &+
    \left(
        21.34_{-0.02}^{+0.02}
    \right),
\end{aligned}
\label{eq:result_radio_sfr_bestfit}
\end{equation}
and shown in Figure~\ref{fig:radio_sfr}. The intrinsic scatter is $\sigma_{\rm int}=0.25_{-0.01}^{+0.01}$~dex, and the Pearson correlation coefficient is $r=0.88$. The fitted slope is consistent with unity within its $1\sigma$ uncertainty. The comparison with radio--SFR relations measured at other frequencies and the physical interpretation of the approximately linear slope are presented in Section~\ref{sec:discussion_radio_sfr}.

\subsection{Mass-dependent semi-calorimetric relation}
\label{sec:results_radio_sfr_semical}

To investigate whether the radio luminosity per unit SFR depends on stellar mass, we fitted the mass-dependent semi-calorimetric model 
\begin{equation}
    L_{\nu,943}
    =
    L_{\rm C}
    \left(
        \frac{\mathrm{SFR}}
             {M_{\odot}\,{\rm yr}^{-1}}
    \right)^a
    \left(
        \frac{M_\star}
             {10^{10}M_{\odot}}
    \right)^{\gamma},
    \label{eq:result_semical_model}
\end{equation}
where $L_{\rm C}$ is the 943~MHz spectral luminosity corresponding to an SFR of $1\,M_\odot\,{\rm yr}^{-1}$ at the reference stellar mass $M_\star=10^{10}M_\odot$, and $\gamma$ describes the stellar-mass dependence of the radio luminosity. In logarithmic form, the relation becomes
\begin{equation}
\begin{split}
    \log_{10}
    \left(
        \frac{L_{\nu,943}}{{\rm W\,Hz^{-1}}}
    \right)
    ={}&
    \log_{10}
    \left(
        \frac{L_{\rm C}}{{\rm W\,Hz^{-1}}}
    \right)
    +
    a\log_{10}
    \left(
        \frac{\mathrm{SFR}}
             {M_{\odot}\,{\rm yr}^{-1}}
    \right)
    \\
    &+
    \gamma
    \log_{10}
    \left(
        \frac{M_\star}
             {10^{10}M_{\odot}}
    \right).
\end{split}
\label{eq:result_semical_logplane}
\end{equation}
Eq.~(\ref{eq:result_semical_logplane}) defines a plane in the three-dimensional $\log L_{\nu,943}$--$\log\mathrm{SFR}$--$\log M_\star$ space. The coefficient $a$ describes the dependence on SFR, $\gamma$ determines the inclination of the plane along the stellar-mass axis, and $\log L_{\rm C}$ sets the normalization. The Bayesian fit gives $\mathrm{log}_{10}[L_C/(\mathrm{W\,Hz^{-1}})]=21.35_{-0.02}^{+0.02}$, $a=0.95^{+0.03}_{-0.03}$, and $\gamma=0.09_{-0.03}^{+0.03}$. The resulting relation is therefore
\begin{equation}
    L_{\nu,943}
    =
    10^{21.35}
    \left(
        \frac{\mathrm{SFR}}
             {M_{\odot}\,{\rm yr}^{-1}}
    \right)^{0.95}
    \left(
        \frac{M_\star}
             {10^{10}M_{\odot}}
    \right)^{0.09}
    {\rm W\,Hz^{-1}}.
    \label{eq:result_semical_bestfit}
\end{equation}
The fit has an intrinsic scatter of $\sigma_{\rm int}=0.25_{-0.01}^{+0.01}$~dex, and a correlation coefficient of $r=0.88$. 

\begin{figure} \includegraphics[width=\columnwidth]{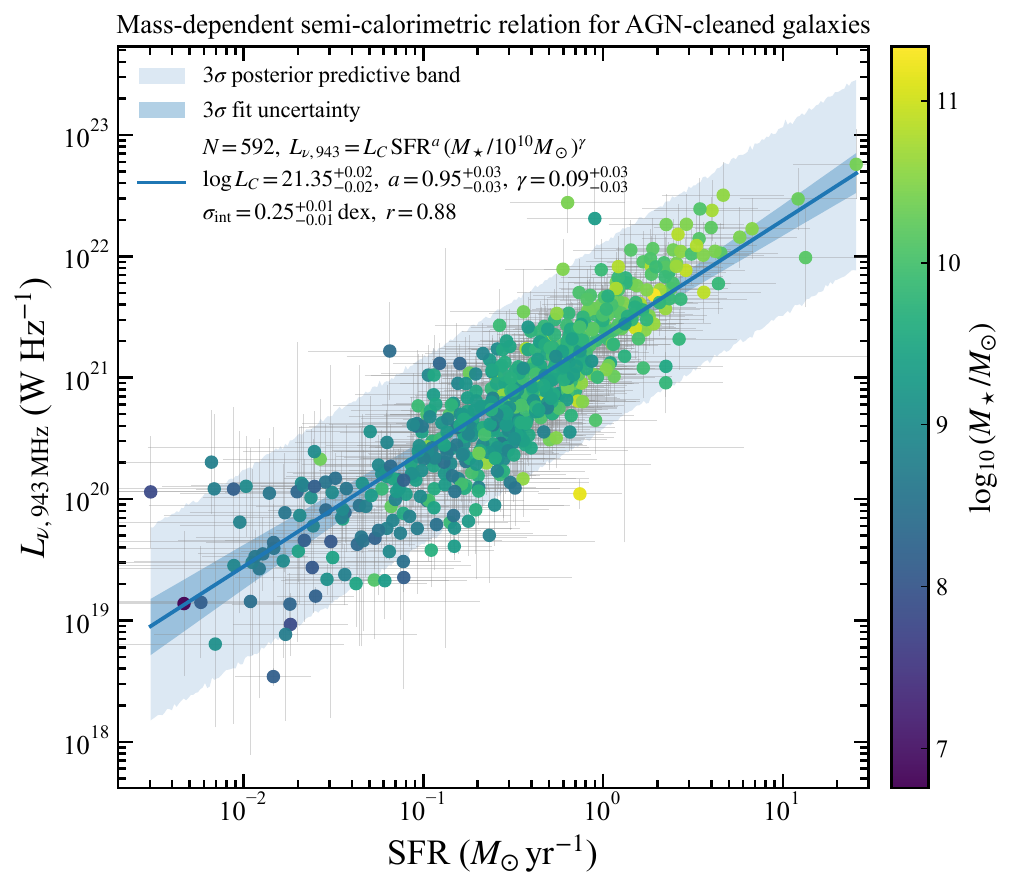}
    \caption{Mass-dependent semi-calorimetric relation between the 943~MHz radio luminosity and star-formation rate for the AGN-cleaned galaxy sample ($N=592$). The data points are color-coded by total stellar mass. The solid blue line represents the median posterior relation evaluated at the reference stellar mass $M_{\star}=10^{10}\,M_{\odot}$. The dark- and light-blue regions have the same meanings as in Figure~\ref{fig:radio_ir_relation}.} 
\label{fig:radio_sfr_semical}
\end{figure}

Figure~\ref{fig:radio_sfr_semical} shows the observed galaxies projected onto the $\log L_{\nu,943}$--$\log\mathrm{SFR}$ plane. The solid blue line corresponds to a slice through the three-dimensional posterior plane at the reference stellar mass $M_\star=10^{10}M_\odot$. Galaxies with different stellar masses are shifted relative to this reference relation according to the fitted mass exponent $\gamma$.

\begin{figure} \includegraphics[width=\columnwidth]{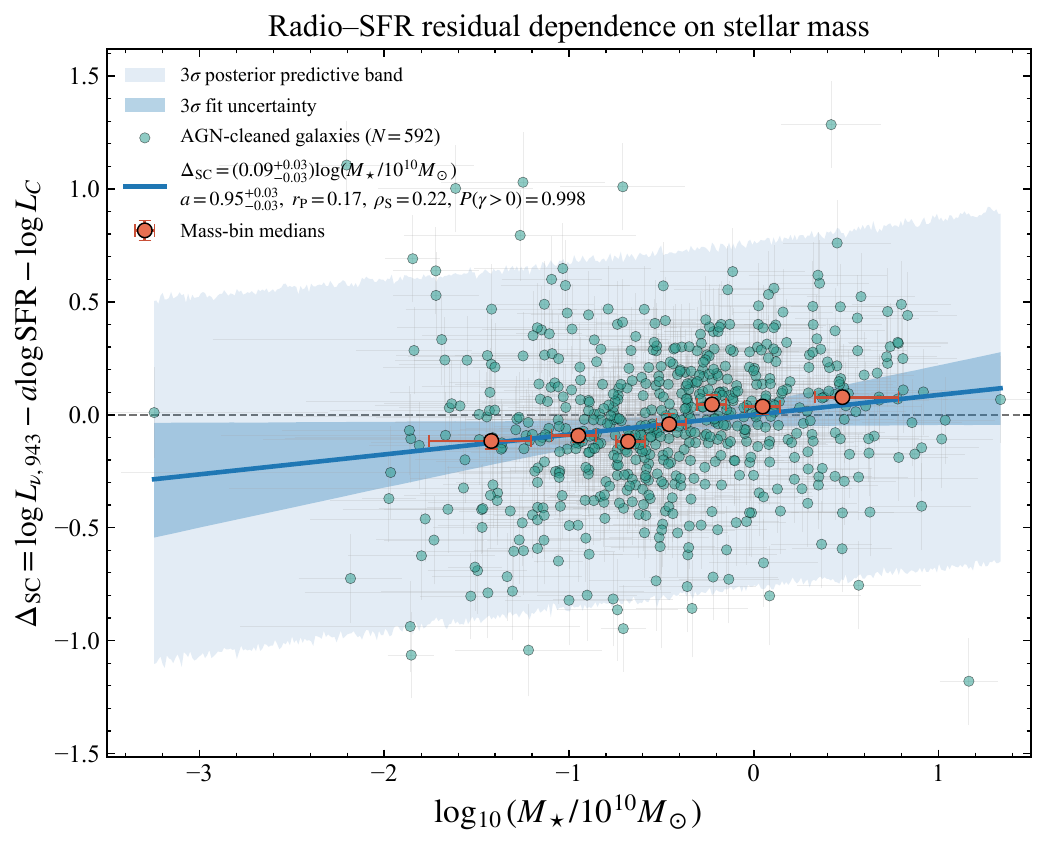}
    \caption{Residual 943~MHz luminosity as a function of stellar mass for the AGN-cleaned sample. Individual galaxies are shown in cyan, while the red points indicate the median residuals in stellar-mass bins. The solid blue line represents the median posterior mass dependence, and the dark and light shaded regions show the $3\sigma$ fit-uncertainty and posterior-predictive intervals, respectively.} 
\label{fig:radio_sfr_residual_mass}
\end{figure}

To visualize the stellar-mass dependence more directly, we removed the fitted SFR dependence and normalization from the radio luminosity and defined the residual as
\begin{equation}
\begin{split}
\Delta_{\rm SC} ={}&
\log_{10}\left(
\frac{L_{\nu,943}}{{\rm W\,Hz^{-1}}}
\right)
-
a\log_{10}\left(
\frac{\mathrm{SFR}}{M_\odot\,{\rm yr}^{-1}}
\right)
\\
&-
\log_{10}\left(
\frac{L_{\rm C}}{{\rm W\,Hz^{-1}}}
\right).
\end{split}
\label{eq:residual_sc}
\end{equation}
Figure~\ref{fig:radio_sfr_residual_mass} shows $\Delta_{\rm SC}$ as a function of stellar mass. The individual galaxies exhibit a weak but positive correlation, with Pearson $r=0.17$ and Spearman $\rho_{\rm S}=0.22$. The binned median residuals and the posterior mass trend both increase with stellar mass, consistent with the positive mass exponent obtained from the Bayesian fit, $\gamma=0.09_{-0.03}^{+0.03}$. The posterior probability of a positive mass dependence is $P(\gamma>0)=0.998$, corresponding to approximately $2.9\sigma$ evidence. The individual residuals show weak positive correlations with stellar mass, consistent with the positive mass exponent favored by the posterior. However, adding stellar mass does not appreciably reduce the intrinsic scatter, which remains approximately $0.25$ dex in both models. The mass term therefore captures a weak systematic dependence at fixed SFR rather than a substantial tightening of the relation. Its physical interpretation is discussed in Section~\ref{sec:discussion_semical}.

\begin{figure} \includegraphics[width=\columnwidth]{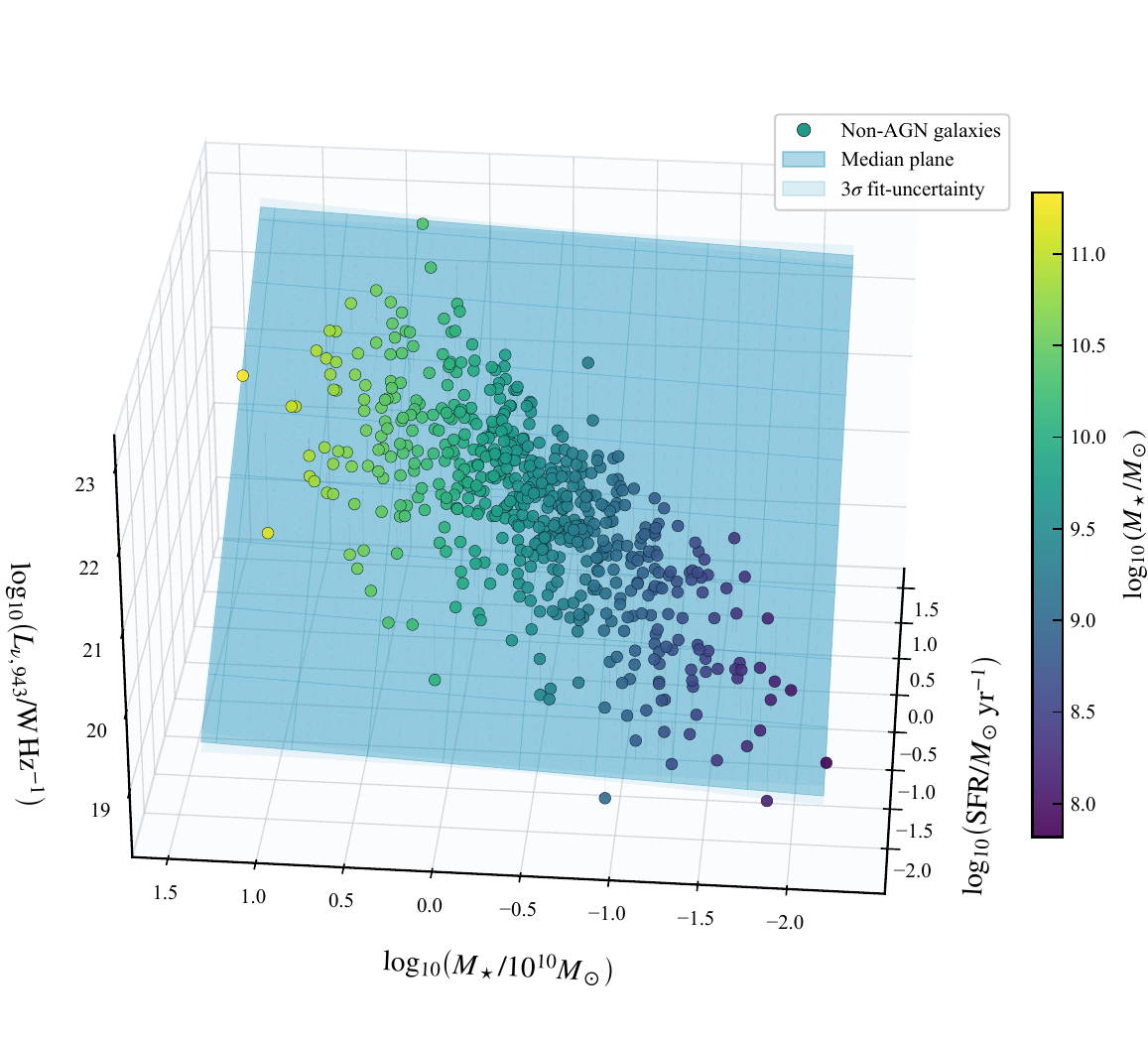}
    \caption{Bayesian semi-calorimetric radio--SFR--stellar-mass relation for the AGN-cleaned sample of 592 galaxies. The data points are color-coded by stellar mass. The blue surface represents the median posterior plane, while the lighter surfaces show the corresponding $3\sigma$ fit-uncertainty envelope. A two-dimensional representation of the best-fitting relation is shown in Figure~\ref{fig:radio_sfr_semical}. } 
\label{fig:radio_sfr_mass_semical}
\end{figure}

\subsection{Rotation-velocity-dependent semi-calorimetric relation}
\label{sec:results_radio_sfr_vrot}

To test whether the radio luminosity depends on the galaxy gravitational potential, we replaced stellar mass with the inclination-corrected maximum rotation velocity and fitted
\begin{equation}
    L_{\nu,943}
    =
    L_{\rm C}
    \left(
        \frac{\mathrm{SFR}}
             {M_\odot\,{\rm yr}^{-1}}
    \right)^a
    \left(
        \frac{V_{\rm rot}}
             {200\,{\rm km\,s^{-1}}}
    \right)^{\delta},
    \label{eq:vrot_semical_model}
\end{equation}
where $L_{\rm C}$ is the normalization at $\mathrm{SFR}=1\,M_\odot\,{\rm yr}^{-1}$ and $V_{\rm rot}=200\,{\rm km\,s^{-1}}$, while $\delta$ describes the rotation-velocity dependence.

The Bayesian fit to the 454 AGN-cleaned galaxies with valid rotation velocities yielded  $\mathrm{log}_{10}[L_C/(\mathrm{W\,Hz^{-1}})]=21.41_{-0.03}^{+0.03}$, $a=1.01_{-0.03}^{+0.03}$, and $\delta=0.24_{-0.08}^{+0.08}$. The fitted relation can therefore be expressed as
\begin{equation}
    L_{\nu,943}
    =
    10^{21.41}
    \left(
        \frac{\mathrm{SFR}}
             {M_\odot\,{\rm yr}^{-1}}
    \right)^{1.01}
    \left(
        \frac{V_{\rm rot}}
             {200\,{\rm km\,s^{-1}}}
    \right)^{0.24}
    {\rm W\,Hz^{-1}}.
\end{equation}
The relation has an intrinsic scatter of $\sigma_{\rm int}=0.24_{-0.01}^{+0.02}$ dex, and a Pearson correlation coefficient of $r=0.88$. The positive value of $\delta$ indicates that, at fixed SFR, galaxies with higher rotation velocities tend to be more radio luminous. 

\begin{figure} \includegraphics[width=\columnwidth]{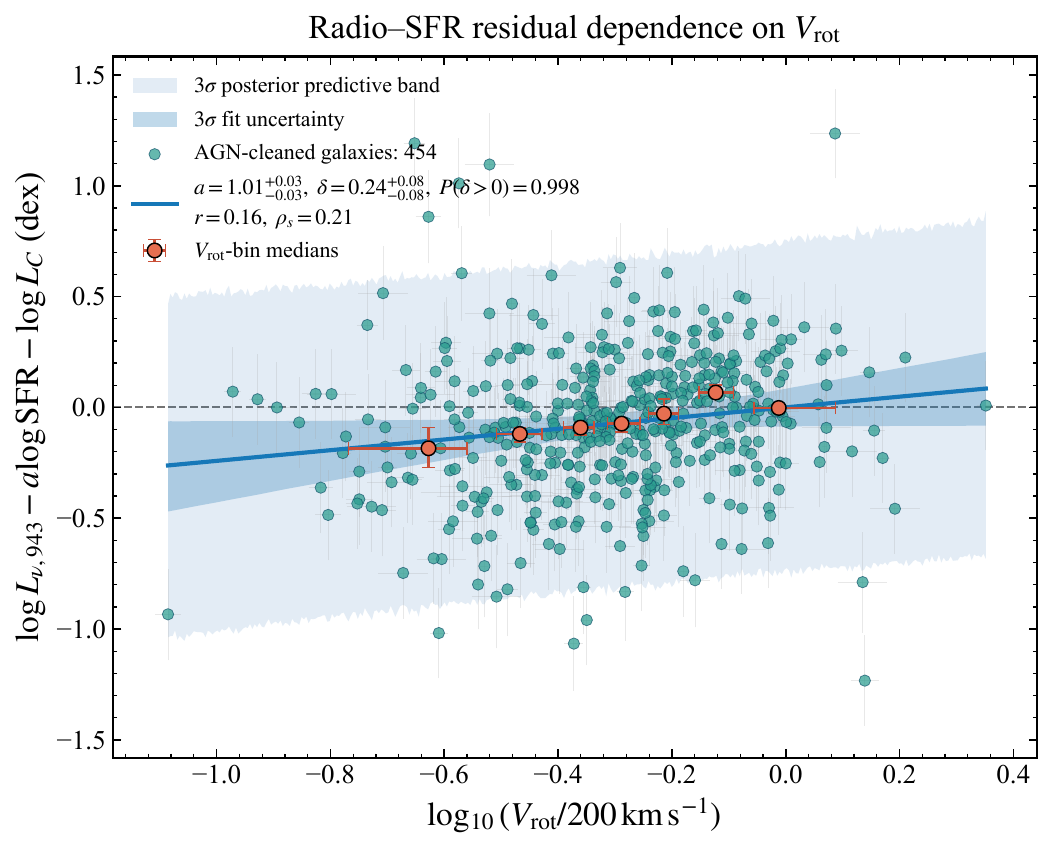}
    \caption{943~MHz radio--SFR residuals as a function of rotation velocity for the AGN-cleaned sample. Individual galaxies are shown in cyan, while the red points indicate median residuals in rotation-velocity bins. The solid blue line represents the median posterior trend. The dark and light shaded regions show the $3\sigma$ fit-uncertainty and posterior-predictive intervals, respectively.} 
\label{fig:radio_sfr_residual_vrot}
\end{figure}

The posterior gives $P(\delta>0)\simeq0.998$, corresponding to approximately $2.9\sigma$ one-sided Gaussian-equivalent evidence for a positive rotation-velocity dependence at fixed SFR. The binned median residuals visualize this weak trend in Figure~\ref{fig:radio_sfr_residual_vrot}. Further discussion of the rotation velocity dependence is presented in Section~\ref{sec:discussion_semical}. The fitted parameters for all scaling relations are summarized in Table~\ref{tab:fit_summary}.

\begin{figure} \includegraphics[width=\columnwidth]{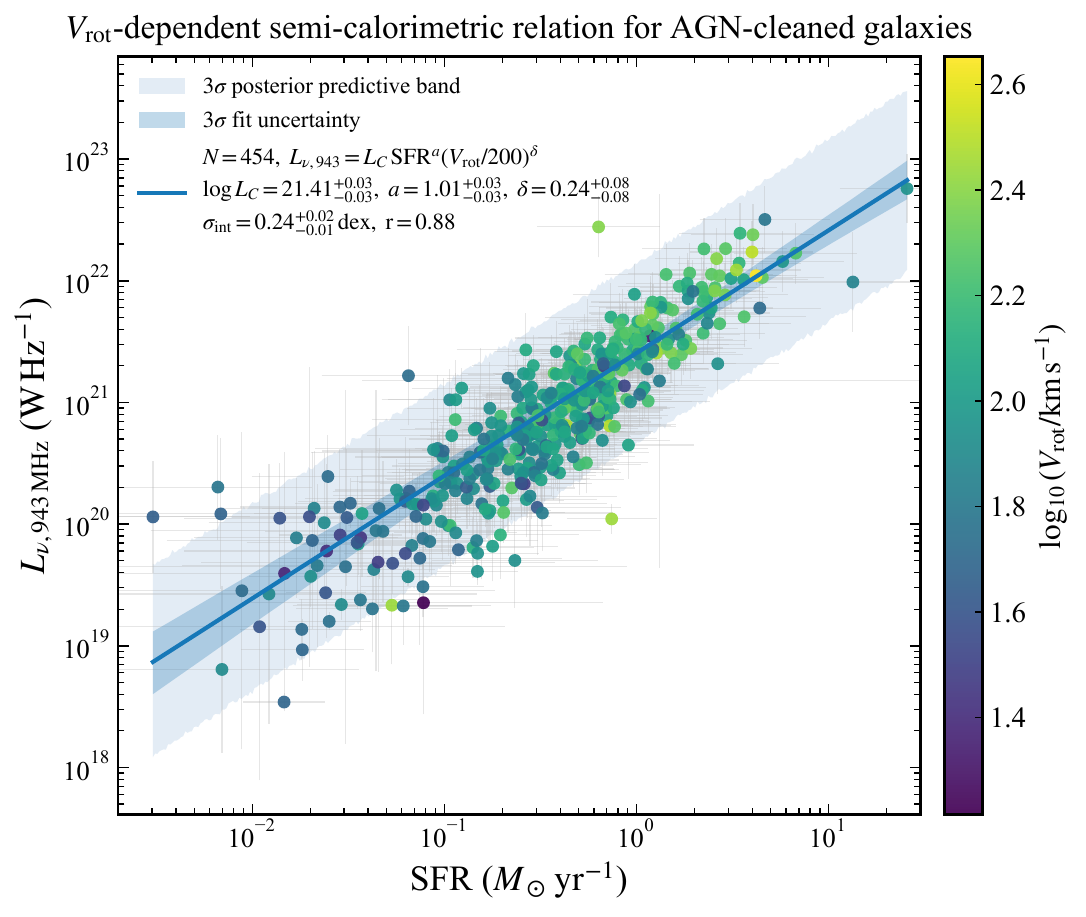}
    \caption{Rotation-speed-dependent radio luminosity–SFR relation for the AGN-cleaned galaxy sample. The 943 MHz radio luminosity is plotted against the SFR for 454 galaxies, with the data points color-coded by the rotational velocity. The solid blue line represents the median posterior relation evaluated at the reference rotation velocity $V_{\rm rot}=200~{\rm km\,s^{-1}}$. The dark- and light-blue regions have the same meanings as in Figure~\ref{fig:radio_ir_relation}.} 
\label{fig:vrot_semicl}
\end{figure}

\begin{deluxetable*}{lcccccc}
\tablecaption{Summary of the fitted scaling relations
\label{tab:fit_summary}}
\tablehead{
\colhead{Relation} &
\colhead{$N$} &
\colhead{$a$} &
\colhead{$\gamma/\delta$} &
\colhead{Normalization} &
\colhead{$\sigma_{\rm int}$} &
\colhead{$r$}
}
\startdata
Radio--W4 full
& 867
& $1.19_{-0.02}^{+0.02}$
& \nodata
& $-5.17_{-0.55}^{+0.52}$
& $0.27_{-0.01}^{+0.01}$
& 0.83
\\
Radio--W4 AGN-cleaned
& 590
& $1.09_{-0.02}^{+0.02}$
& \nodata
& $-2.97_{-0.49}^{+0.48}$
& $0.21_{-0.01}^{+0.01}$
& 0.89
\\
Radio--W4 AGN-host
& 277
& $1.43_{-0.07}^{+0.07}$
& \nodata
& $-10.63_{-1.63}^{+1.48}$
& $0.36_{-0.02}^{+0.02}$
& 0.74
\\
Radio--SFR
& 592
& $1.02_{-0.02}^{+0.02}$
& \nodata
& $21.34_{-0.02}^{+0.02}$
& $0.25_{-0.01}^{+0.01}$
& 0.88
\\
Radio--SFR+$M_\star$
& 592
& $0.95_{-0.03}^{+0.03}$
& $\gamma=0.09_{-0.03}^{+0.03}$
& $21.35_{-0.02}^{+0.02}$
& $0.25_{-0.01}^{+0.01}$
& 0.88
\\
Radio--SFR+$V_{\rm rot}$
& 454
& $1.01_{-0.03}^{+0.03}$
& $\delta=0.24_{-0.08}^{+0.08}$
& $21.41_{-0.03}^{+0.03}$
& $0.24_{-0.01}^{+0.02}$
& 0.88
\\
\enddata

\tablecomments{
The normalization is the intercept $b$ for the radio--W4 relations, the logarithmic radio luminosity at $\mathrm{SFR}=1~M_\odot\,{\rm yr}^{-1}$ for the radio--SFR relation, and $\log_{10}(L_{\rm C}/{\rm W\,Hz^{-1}})$ for the mass- and rotation-velocity-dependent models. For the radio--W4 relations, the quoted scatter is orthogonal; all other scatter values refer to radio luminosity at fixed predictors. All scatter values are in dex, and $r$ is the Pearson correlation coefficient.
}
\end{deluxetable*}

\section{Discussion}
\label{sec:discussion}

\subsection{The radio--infrared luminosity relation}
\label{sec:discussion_radio_ir}

The close radio--W4 correlation is expected because both quantities ultimately trace recent star formation, although variations in dust heating and CRE transport can modify its slope and scatter. For the AGN-cleaned sample, the Bayesian orthogonal regression gives $L_{\nu,943}\propto L_{\nu,W4}^{1.09}$. The relation is therefore close to linear but remains mildly superlinear, indicating that the radio luminosity per unit W4 luminosity increases weakly toward more infrared-luminous galaxies. This trend is qualitatively consistent with the superlinear low-frequency radio--infrared relation reported by~\citet{Wang2019}, although their slope of approximately $1.37$ at 150~MHz is steeper than that measured here at 943~MHz. The difference may partly reflect the greater sensitivity of lower-frequency synchrotron emission to long-lived CREs, propagation, and escape. 

The mild superlinearity and remaining scatter may reflect variations in both W4 emission and synchrotron-radiation efficiency. W4 luminosity depends on dust abundance, metallicity, temperature, and geometry. Dust deficiency can suppress W4 emission at the low-luminosity end, tending to flatten the radio--W4 relation, all else being equal. In dusty, highly inclined galaxies, internal dust absorption may reduce the observed W4 emission~\citep{Vargas2018}. If this suppression preferentially affects the high-luminosity end, it could contribute to the observed mild superlinearity. Dust heating by older stellar populations can also modify W4 emission~\citep{Kennicutt2009,Cluver2014}, while variations in magnetic-field strength and CRE confinement can alter synchrotron efficiency. These effects are particularly relevant for a heterogeneous nearby-galaxy sample spanning dwarf galaxies, normal spirals, and a broad range of star-formation environments \citep{Leroy2008}. The present data do not establish which effect dominates. Matched far-infrared measurements and spatially resolved, multi-frequency radio observations would help distinguish dust-related effects on W4 emission from variations in CRE transport and synchrotron efficiency.

The AGN-host relation is considerably steeper and more dispersed than the AGN-cleaned relation. Nuclear radio emission can increase $L_{\nu,943}$ without a corresponding increase in W4 emission from star formation, while AGN-heated dust can contribute differently among AGN populations. The full-sample relation consequently lies between the AGN-cleaned and AGN-host results. This demonstrates that mixing star-formation-dominated galaxies with AGN hosts alters both the slope and scatter of the integrated radio--infrared relation.

Finally, the AGN-cleaned sample contains galaxies without an \emph{identified} AGN and should not be interpreted as a demonstrably pure star-forming sample. External AGN catalogs, SIMBAD classifications, and WISE color diagnostics have different selection functions and may miss weak, obscured, or host-dominated nuclei. Residual AGN activity may therefore contribute to the remaining scatter, although the tighter AGN-cleaned relation indicates that the adopted screening removes a substantial fraction of the contamination.

\subsection{The approximately linear 943~MHz radio--SFR relation}
\label{sec:discussion_radio_sfr}

The fitted 943~MHz radio--SFR slope, $a=1.02_{-0.02}^{+0.02}$, is consistent with unity. In the ideal electron-calorimeter limit, where CREs lose their energy before escaping, radio luminosity is expected to scale near-unity slope with SFR, provided that the CRE injection efficiency and the fraction of energy radiated through synchrotron emission remain similar among galaxies~\citep{Voelk1989,Lacki2010a}. However, exact linearity is not required at a fixed observing frequency. As magnetic-field strength increases with SFR, the observed emission probes lower-energy CREs. For an injected electron spectrum $N(E)\propto E^{-p}$, where $N(E)$ denotes the injected CRE number distribution per unit energy, this magnetic-field effect can produce mild superlinearity even in the calorimetric limit when $p>2$~\citep{Pfrommer2022}. Our approximately linear relation is therefore  consistent with a calorimetric interpretation, but its slope alone does not demonstrate complete calorimetry, because systematic variations in CRE injection, magnetic-field strength, energy losses, and escape may partially compensate for one another.

The intrinsic scatter of approximately 0.25~dex indicates that SFR alone does not fully determine the radio luminosity. Galaxy-to-galaxy variations in magnetic fields, CRE losses, transport, and escape can contribute to this dispersion. Possible secondary dependences on stellar mass and rotation velocity are discussed in Section~\ref{sec:discussion_semical}.

Our result at 943~MHz is close to several large-sample 150~MHz results, including $1.07\pm0.01$ from~\citet{Gurkan2018}, $1.058\pm0.007$ from~\citet{Smith2021}, $1.08\pm0.06$ from~\citet{Best2023}, and $1.019\pm0.009$ from~\citet{Soumyadeep2024}. However, steeper relations have also been reported:~\citet{Brown2017} obtained $1.16\pm0.05$ at 150~MHz, while~\citet{Wang2019} and \citet{Heesen2022} found strongly superlinear relations at 144--150~MHz. At GHz frequencies, \citet{Heesen2014} measured $1.11\pm0.08$ at 1.4~GHz, whereas~\citet{Li2016} obtained $1.13\pm0.07$ at 1.6~GHz and $1.06\pm0.08$ at 6~GHz. Our result therefore does not follow a simple monotonic sequence between the low- and high-frequency measurements.

Observing frequency alone cannot account for these differences. The measured slope is also sensitive to sample selection, radio-detection limits, SFR estimation, stellar-mass distribution, and regression methodology~\citep{Davies2017,Smith2021,Bonato2021,Soumyadeep2024}. In our radio-detected sample, low-SFR galaxies are preferentially retained when they are relatively radio bright, elevating the low-SFR end and potentially flattening the fitted relation. Different SFR tracers may introduce additional systematic differences, particularly for low-mass, dust-poor galaxies whose mid-infrared emission may not recover the full unobscured star formation. Galaxy-to-galaxy variations in the radio spectral index, magnetic field, and CRE transport can further modify how measurements at different frequencies trace the underlying SFR~\citep{Heesen2022}. A robust test of frequency dependence therefore requires the same galaxy sample to be analyzed at multiple frequencies with consistent sensitivity limits, SFR tracers, and regression methods.

The integrated relation averages over CRE injection, propagation, and energy losses throughout each galaxy. Spatially resolved observations provide a complementary way to separate these processes. CRE propagation can redistribute radio emission from regions of high $\Sigma_{\rm SFR}$ into the surrounding disk and halo, producing sublinear resolved relations even when the integrated relation is approximately linear~\citep{Heesen2014}. \citet{Heesen2024} measured resolved slopes of $0.86\pm0.02$ at 144~MHz and $0.92\pm0.01$ at 1365~MHz, which approached unity when young CREs were selected using the radio spectral index. Similarly, \citet{Xu2026} found that the median resolved nonthermal radio--SFR slope in 19 edge-on galaxies increases from approximately $0.57$ at 144~MHz to $0.95$ at 6~GHz. They interpreted this trend in terms of CRE cooling and transport, with line-of-sight projection further modifying the resolved relation. Future resolved, multi-frequency observations of our sample will help distinguish variations in galaxy-scale CRE confinement from CRE transport and aging within individual galaxies.

\subsection{Mass and rotation-velocity dependence}
\label{sec:discussion_semical}

The mass-dependent semi-calorimetric fit presented in
Section~\ref{sec:results_radio_sfr_semical} yields $\log_{10}(L_{\rm C}/{\rm W\,Hz^{-1}})=21.35_{-0.02}^{+0.02}$ and $\gamma=0.09_{-0.03}^{+0.03}$. The positive mass exponent indicates that, at fixed SFR, more massive galaxies tend to be slightly more radio luminous. An increase of one decade in mass produces an increase in radio luminosity by a factor $10^{\gamma}=10^{0.09}\simeq1.23$. 

Positive secondary stellar-mass dependences have been reported in several previous galaxy-scale studies. At 144--150~MHz, mass exponents of $\gamma \sim 0.2$--$0.4$ have been measured using both empirical and semi-calorimetric approaches~\citep{Gurkan2018,Smith2021,Soumyadeep2024,Heesen2022}. More recently, \citet{Heesen2026} obtained $a=1.10\pm0.06$ and $\gamma=0.23\pm0.05$ at 144~MHz. Similar stellar-mass trends have also been found in studies of the radio--SFR relation, the infrared--radio correlation, and multivariate analyses of star-forming galaxies~\citep{Bonato2021,Delvecchio2021,Shenoy2026}. These results provide consistent evidence for a secondary stellar-mass dependence, although the fitted exponents are not directly comparable because of differences in observing frequency, model parameterization, sample selection, and SFR estimation.

In the semi-calorimetric interpretation, stellar mass is an empirical proxy for CRE confinement and radiative efficiency rather than a direct measurement of either quantity \citep{Heesen2022}. More massive galaxies are generally larger and may possess stronger or more extended magnetic fields and longer CRE residence times. These properties can reduce CRE escape and increase the fraction of particle energy radiated through synchrotron emission before the CREs leave the galaxy~\citep{Lacki2010,Heesen2022}. Stellar mass also correlates with gas content, gravitational potential, and outflow properties, which may further influence CRE transport. The observed mass dependence therefore probably represents the combined influence of several correlated galaxy properties rather than a single physical mechanism.

Numerically, the mass dependence measured at 943~MHz is weaker than those generally reported at 144--150~MHz. This difference may partly reflect the observing frequency. The lower-energy CREs dominating the 144~MHz emission can remain observable for longer and propagate farther from their acceleration sites, making the integrated emission more sensitive to galaxy size, confinement, and escape. Selection effects may also contribute: requiring a 943~MHz detection preferentially removes radio-faint, low-mass galaxies and may reduce the observed contrast between the low- and high-mass populations. Differences in the sampled mass and SFR ranges, parameter estimates, AGN rejection, and galaxy morphology may introduce additional variation among studies.

As a complementary test, we replaced stellar mass with the inclination-corrected maximum rotation velocity. The fit presented in Section~\ref{sec:results_radio_sfr_vrot} yields $\log_{10}(L_{\rm C}/{\rm W\,Hz^{-1}})=21.41_{-0.03}^{+0.03}$ and $\delta=0.24_{-0.08}^{+0.08}$. Unlike stellar mass, $V_{\rm rot}$ provides a kinematic proxy for the depth of the gravitational potential. Using the stellar-mass Tully--Fisher relation $M_\star\propto V_{\rm rot}^{3.60\pm0.13}$~\citep{Reyes2011}, the measured mass dependence predicts $\delta_{\rm pred}=3.60\gamma=0.32\pm0.11$, consistent with the directly fitted value of $\delta=0.24\pm0.08$. This agreement suggests that the stellar-mass and rotation-velocity trends may trace a common underlying dependence on galaxy structure and CRE confinement. 

The spectral-index-dependent relation of~\citet{Heesen2026}
provides a complementary comparison. Combining their Equation~(5) with the correlations in their Table~1 and assuming a power-law radio spectrum gives expected exponents of $\gamma_{\rm pred}\simeq0.12$ and $\delta_{\rm pred}\simeq0.39$ at 943~MHz. Our measured mass dependence agrees closely with this expectation, while the rotation-velocity dependence is weaker but remains consistent within the combined uncertainties. Consistent with this interpretation, \citet{Heesen2022} found that rotation velocity correlates with the integrated radio spectral index, with faster-rotating galaxies tending to exhibit steeper spectra. This provides evidence that the relation between $V_{\rm rot}$ and radio efficiency may be mediated by CRE aging and confinement rather than by gravity alone.

The radio spectral index provides a more direct diagnostic of CRE aging, radiative losses, and escape. \citet{Heesen2024} recovered an approximately linear radio--SFR relation after correcting for a spectral-index-dependent calorimetric efficiency, while~\citet{Heesen2026} reported a plane involving radio luminosity, SFR, and spectral index. Other galaxy properties, including physical size, SFR surface density, gas surface density, and magnetic-field strength, have also been linked to radio--SFR residuals and CRE transport~\citep{Heesen2019,Heesen2022,Heesen2024,Niklas1997,Lacki2010, Chyzy2017,Heesen2023}. The existence of a plane alone, however, does not establish a fundamental physical relation. In particular, adding stellar mass does not substantially reduce the intrinsic scatter relative to the radio--SFR relation, while the rotation-velocity fit applies to a smaller subsample and therefore cannot be compared directly in terms of scatter. The detected secondary trends should thus be interpreted primarily as empirical correlations rather than as evidence that either parameter uniquely controls the radio luminosity.

\section{Conclusions}
\label{sec:conclusions}

We investigated the integrated 943~MHz radio--infrared and radio--SFR relations of 875 nearby galaxies within $D<50$~Mpc using ASKAP/EMU observations combined with UV and infrared measurements. Our main conclusions are as follows.

\begin{enumerate}[
    leftmargin=*,
    label=\arabic*.,
    itemsep=0pt,
    parsep=0pt,
    topsep=0pt
]

\item The 943~MHz radio luminosity is strongly correlated with the WISE W4 luminosity. The AGN-cleaned sample follows a nearly linear and relatively tight relation, whereas the AGN-host relation is steeper and more dispersed. This difference indicates that varying contributions from nuclear radio emission and AGN-heated dust modify both the slope and scatter of the integrated radio--infrared relation.

\item The integrated 943~MHz radio--SFR relation is approximately linear, with a slope of $1.02\pm0.02$ and an intrinsic scatter of $0.25\pm0.01$ dex. Although this mean relation is consistent with the electron-calorimeter expectation, a linear slope alone does not demonstrate complete calorimetry. The remaining scatter indicates that CRE confinement, magnetic fields, and synchrotron-radiation efficiency vary among galaxies.

\item The semi-calorimetric model reveals a weak but positive secondary stellar-mass dependence, with $\gamma=0.09\pm0.03$. At fixed SFR, more massive galaxies tend to be slightly more radio luminous, suggesting that stellar mass traces galaxy properties that regulate CRE confinement and synchrotron-radiation efficiency.

\item A complementary fit reveals a positive rotation-velocity dependence, with $\delta=0.24\pm0.08$. This result is consistent with the exponent $\delta_{\rm pred}=0.32\pm0.11$ predicted from the measured stellar-mass dependence and the stellar-mass Tully--Fisher relation. The agreement suggests that the stellar-mass and rotation-velocity trends may both reflect the influence of the galaxy gravitational potential on CRE confinement and synchrotron-radiation efficiency.

\end{enumerate}
Previous multi-frequency and spatially resolved studies have established that CRE aging, transport, and escape affect the radio--SFR relation~\citep{Heesen_2018b,Heesen2024,Heesen2026,Xu2026}, but their relative contributions to the galaxy-integrated scatter remain uncertain. Matched-resolution observations with SKA-Low and SKA-Mid would enable spatially resolved measurements of spectral indices and curvature while recovering faint emission from extended radio halos. Combined with matched SFR maps, these observations could separate the effects of magnetic-field amplification, radiative losses, and energy-dependent CRE escape in star-forming disks and halos.

\begin{acknowledgments}
We thank the reviewer for constructive comments that have significantly improved this paper. This research has been supported by the National SKA Program of China (2022SKA0120101). J.T.L. acknowledges the financial support from the National Natural Science Foundation of China (NSFC) through the grants 12321003 and 12273111, the science research grants from the China Manned Space Program with grant no. CMS-CSST-2025-A04 and CMS-CSST-2025-A10, and Jiangsu Innovation and Entrepreneurship Talent Team Program through the grant JSSCTD202436. This scientific work uses data obtained from Inyarrimanha Ilgari Bundara / the Murchison Radio-astronomy Observatory. We acknowledge the Wajarri Yamaji People as the Traditional Owners and native title holders of the Observatory site. The Australian SKA Pathfinder is part of the Australia Telescope National Facility, which is managed by CSIRO. Operation of ASKAP is funded by the Australian Government with support from the National Collaborative Research Infrastructure Strategy. Establishment of ASKAP, the Murchison Radio-astronomy Observatory and the Pawsey Supercomputing Centre are initiatives of the Australian Government, with support from the Government of Western Australia and the Science and Industry Endowment Fund.
\end{acknowledgments}

\facilities{ASKAP, GALEX, WISE}

\software{ASKAPsoft~\citep{guzman2019}, Astropy~\citep{astropy2013,astropy2018,astropy2022}, emcee~\citep{ForemanMackey2013}}

\bibliography{radio_ir}{}
\bibliographystyle{aasjournalv7}
\end{document}